\documentclass[12pt]{article}
\usepackage{amsfonts,amssymb,epsfig,amsmath}
\usepackage{color}
\renewcommand{\text}[1]{#1}

\newcommand{\be}{\begin{equation}}
\newcommand{\ee}{\end{equation}}
\newcommand{\ben}{\begin{displaymath}}
\newcommand{\een}{\end{displaymath}}
\newcommand{\bea}{\begin{eqnarray}}
\newcommand{\eea}{\end{eqnarray}}
\newcommand{\bean}{\begin{eqnarray*}}
\newcommand{\eean}{\end{eqnarray*}}
\newcommand{\nn}{\nonumber \\}
\newcommand{\ba}{\begin{array}}
\newcommand{\ea}{\end{array}}
\newcommand{\bi}{\begin{itemize}}
\newcommand{\ei}{\end{itemize}}

\renewcommand{\theequation}{\arabic{section}.\arabic{equation}}
\def\theequation{\thesection.\arabic{equation}}

\def\l{\lambda}
\def\a{\alpha}

\def\b{\beta}
\def\g{\gamma}
\def\G{\Gamma}

\def\G{\Gamma}
\def\g{\gamma}
\def\e{\epsilon}
\def\s{\sigma}
\def\e{\epsilon}

\DeclareMathOperator{\vol}{vol}

\begin{document}

\makeatletter
\renewcommand{\theequation}{\thesection.\arabic{equation}}
\@addtoreset{equation}{section}
\makeatother

\begin{titlepage}

\vfill
\begin{flushright}
KIAS-P10012
\end{flushright}

\vfill

\begin{center}
   \baselineskip=16pt
   {\Large\bf Supersymmetric $AdS_3 \times S^2$ M-theory geometries with fluxes. }
   \vskip 2cm
     Eoin \'O Colg\'ain$^{\clubsuit}$, Jun-Bao Wu$^{\clubsuit}$ \& Hossein Yavartanoo$^{\spadesuit}$
       \vskip .6cm
             \begin{small}
      \textit{Korea Institute for Advanced Study$^{\clubsuit}$, \\
        Seoul 130-722, Korea \\
        Department of Physics, Kyung Hee University$^{\spadesuit}$, \\Seoul 130-701, Korea}
        \end{small}\\*[.6cm]
\end{center}

\vfill
\begin{center}
\textbf{Abstract}\end{center}

\begin{quote}
Motivated by a recent observation that the LLM geometries admit 1/4-BPS M5-brane probes with worldvolume $AdS_3 \times \Sigma_2 \times S^1$ preserving the R-symmetry, $SU(2)\times U(1)$, we initiate a classification of the most general $AdS_3 \times S^2 $ geometries in M-theory dual to two-dimensional chiral $\mathcal{N}= (4,0)$ SCFTs. We retain all field strengths consistent with symmetry and derive the torsion conditions for the internal six-manifold, $M_6$, in terms of two linearly independent spinors. Surprisingly, we identify three Killing directions for $M_6$, but only two of these generate isometries of the overall ansatz. We show that the existence of this third direction depends on the norm of the spinors. With the torsion conditions derived, we establish the MSW solution as the only solution in the class where $M_6$ is an $SU(3)$-structure manifold. Then, specialising to the case where the spinors define an $SU(2)$-structure, we note that supersymmetry dictates that all magnetic fluxes necessarily thread the $S^2$.  Finally, by assuming that the two remaining Killing directions are parallel and aligned with one of the two vectors defining the $SU(2)$-structure, we derive a general relationship for the two spinors before extracting a known class of solutions from the torsion conditions.
\end{quote}
\vfill

\end{titlepage}

\section{Introduction}
This work touches on two beautiful and exciting developments in the string theory literature over the last few years. Of most recent interest is a new interpretation of a class of 4d theories arising as the infra-red fixed point of compactification of a 6d superconformal theory on a Riemann
surface $\Sigma_2$ with punctures \cite{Gaiotto:2009we}.  This paved the way for the elegant observation that the Nekrasov instanton partition function in 4d is identical to the conformal block in a 2d
Liouville CFT \cite{AGT} and has led to a flurry of studies focussing on non-local operators in these theories \cite{nonlocalops,Chen:2010jg}. Secondly, this work also revisits supersymmetric bubbling geometries in string and M-theory. Starting with the seminal work of \cite{LLM}, we have witnessed the identification of a host of solutions preserving different degrees of supersymmetry \cite{Donos1, eoin, bubbles}. A useful overview of developments may be found in \cite{Chen2}.


The holographic dual of the theories of \cite{Gaiotto:2009we} has been identified in \cite{MG}, where supersymmetry preserving M5-branes with worldvolume $AdS_5 \times S^1$,  were considered in the LLM geometries (including the Maldacena-N\'u\~nez (MN) geometry \cite{MN}) dual to $\mathcal{N}=2$ SCFTs with R-symmetry $SU(2) \times U(1)$ \cite{LLM}.  Following a study of supersymmetric M5-brane probes \cite{Chen:2010jg} in the MN \cite{MN}  background, it has been noted that the geometry also permits a 1/4-BPS M5-brane surface operator with worldvolume $AdS_3 \times H^2 \times S^1$, where $H^2$ denotes the hyperbolic space of MN. In addition, the $SU(2) \times U(1)$ R-symmetry is preserved\footnote{As MN is one simple case in a class of LLM geometries sharing similar structure, all these M5-brane probes also exist in the other LLM geometries. }. Within this context, the primary goal of this work is to initiate a classification of 1/4-BPS solutions of M-theory of the form $AdS_3 \times S^2$, where the two-sphere ansatz corresponds to the $SU(2)$ R-symmetry of the dual SCFT, with the sideline objective of finding the back-reacted geometry corresponding to the M5-probe of \cite{Chen:2010jg}. See \cite{backx} for other studies on the back-reaction of non-local operators.

We approach this problem by employing G-structures (for selected other works in 11d, see \cite{jerome1,Gstructure,d11kill,oisin}) following a two-stage decomposition of the 11d Killing spinor equation (KSE) pioneered in LLM \cite{LLM}. As we will see, this procedure leads to differential and algebraic Killing spinor equations in 6d in terms of two independent, non-chiral spinors. In a departure from previous work in this direction, we make neither a simplifying assumption about the fluxes from the offset as in \cite{nakwoo}\footnote{We remark that the gamma matrix decomposition utilised in \cite{nakwoo} only permits a limited flux ansatz and attempts to incorporate more fluxes necessitate a return to an LLM-type decomposition.}, nor after reducing the Killing spinor equation as in \cite{LLM}. Instead, after deriving the most general torsion conditions for this decomposition in terms of spinor bilinears, we assume chiral $\mathcal{N} = (4,0)$ supersymmetry in the dual 2d SCFT and in turn consider an internal $M_6$ with both $SU(3)$-structure and $SU(2)$-structure. The expected 8 supercharges preserved by these geometries arise from the reduction ansatz; an $SU(2)$-doublet from the two-sphere and two Poincar\'e and two superconformal supercharges from the $AdS_3$ spacetime. The counting depends on the proviso that one imposes enough projection conditions internally so that the 6d Killing spinor has a single real component. Indeed, the supersymmetry conditions we derive later show perfect agreement with the work of \cite{nakwoo}, which in turn has been shown to be consistent to \cite{wrappedbranes}\footnote{Recall that instead of using G-structures as a classification tool, this paper determines the geometric constraints on supersymmetric solutions arising as $AdS$ limits from geometries admitting wrapped branes.}. The class of solutions presented in these papers all preserve chiral $\mathcal{N} = (4,0)$.

In general, this work is intended to be a stepping stone towards a complete classification of M-theory geometries with $AdS_3 \times S^2$ factors, with the determination of the general torsion conditions. The results for $M_6$ admitting $SU(3)$ and $SU(2)$-structure we present here, while leaving the generalisation to smaller structure groups for future work. Remarkably, if one assumes $M_6$ permits only a single chiral spinor defining $SU(3)$-structure, then the constraints derived from the torsion conditions are enough to identify \cite{MSW} as the \textit{only} solution. Then, when $M_6$ admits two orthogonal, chiral spinors defining an $SU(2)$-structure, supersymmetry alone tells us that there is no four-form flux on $M_6$, while the torsion condition constraints, in tandem with the added assumption that the Killing direction is along only one of the two $SU(2)$-structure vectors, are enough so that we can uniquely determine the relationship between the two original spinors. One may remember that \cite{LLM} assumed that the two spinors were linearly dependent, so this work attempts to see to what extent that choice is motivated by supersymmetry. Therefore, our analysis when $M_6$ is an $SU(2)$-structure manifold, provides a strong case for the class of solutions identified in \cite{nakwoo,wrappedbranes} to be the only class of solutions with $M_6$ admitting $SU(2)$-structure. Unfortunately, we are unaware of any explicit, regular solutions in this class, however such a discovery would open up the exciting prospect of analysing properties of the dual SCFT.

Having sketched the landscape of the setting for our work, we pause to comment on the secondary motivation of back-reacting the 1/4-BPS M5-brane probes of \cite{Chen:2010jg}. On the requirement that the $SU(2) \times U(1)$ R-symmetry is preserved, early strong candidates for a back-reacted solution may already be found in the literature. Among the solutions of Romans' 5d $SU(2) \times U(1)$ gauged supergravity \cite{romans} uplifted to LLM in \cite{oscar}, the magnetovac solutions are most promising as the $U(1)$-fibration is the same as LLM, however the $SU(2)$-symmetry is broken to $U(1)$. The class of solutions presented in \cite{nakwoo,wrappedbranes} have the obvious draw-back that it is difficult to find explicit solutions. Our hope at the beginning of this project was that the inclusion of a more general flux ansatz may lead to a more favourable $U(1)$-fibration and a clean one-parameter family of solutions as in the similar set-up of \cite{jerome1}. Our results indicate this is not the case for $M_6$ being an $SU(3)$ or $SU(2)$-structure manifold, however we cannot rule out this possibility with smaller structure groups.

The format for the rest of the paper runs as follows. Section 2 concerns itself with the decomposition of the Killing spinor equation. In general, one is left with two linearly independent, non-chiral spinors $\e_+,\e_-$, but they will be related once we specify the structure group (see \cite{jerome5}). Section 3 enumerates the conditions arising from the general torsion conditions and the algebraic constraints. Interestingly, demanding consistency amongst the torsion conditions leads to a single expression that may not be derived from the algebraic constraints. In section 4 we impose $SU(3)$ and $SU(2)$-structure on the internal $M_6$ and identify how $\e_+, \e_-$ may be rewritten in terms of unit norm, chiral spinors. In each case, we examine the relationship between $\e_{+}$ and $\e_{-}$ in the light of the constraints derived in section 3. In section 5 we derive the most general solution consistent with $SU(2)$-structure and comment on properties of this class of solutions.

\section{Killing spinor decomposition}
As stated in the introduction, the focus of this paper is $AdS_3$ M-theory solutions preserving $SU(2)$ superconformal symmetry where the $SU(2)$ symmetry corresponds to the isometries of an internal, round $S^2$ \footnote{It is possible to consider a slightly more general ansatz with $S^2$ by considering the Killing direction $\psi$ to be fibred over the $S^2$, however it is shown in \cite{LLM} that the resulting supercharges do not form a doublet of $SU(2)$.} on the transverse eight-dimensional space $M_8$. In contrast to the work of \cite{nakwoo}, where the restricted flux ansatz permitted a gamma matrix decomposition in terms of a direct product of spinors, in order to retain more fluxes, one can adopt a two stage decomposition \cite{LLM} (see \cite{jerome1,dario,Gauntlett:2005ww} for simpler reductions). The first step will involve reduction on $AdS_3$, followed by a second step reduction on $S^2$.

As mentioned in the introduction, it is expected, with suitable internal projectors on the internal $M_6$, that the dual SCFTs correspond to chiral $\mathcal{N} = (4,0)$. The eight preserved supercharges can also be accounted for in the analytically continued $S^3 \times S^2$ geometry (see \cite{nakwoo}) where they come in irreducible representations of the isometry group i.e. an $SU(2)$-doublet and $(2,1) \oplus (1,2)$ of $SU(2) \times SU(2) \subset SO(4)$. The fact that the chirality of the Killing spinors is the same may be deduced from the explicit expressions for $AdS$ Killing spinors presented in \cite{pope1}. One simply has to take the large $AdS$ radius, in which case the superconformal Killing spinors which initially have mixed chirality under $\tau_{01}$, become chiral, matching the chirality of the Poincar\'e Killing spinors.

We begin the procedure with the 11d gamma matrices which may be decomposed in a $(3,8)$-split as:
\bea
\G_{a} &=& \tau_a \otimes \g_9, \nn
\G_{m} &=& 1 \otimes \g_{m},
\eea
with indices $a=0,1,2$ and $m = 1,...,8$. The $\tau$ matrices are simply related to the Pauli matrices $ \tau_0 = i \sigma_3, \tau_{i} = \sigma_i$ and choosing $\G_{0123456789\sharp} = -1$ means that $\g_{9} = \g_{12345678}$, with $\g_9^2 = 1$. The metric may then be dimensionally reduced via the ansatz
\bea
ds^{2} &=& e^{2 \lambda} \left[  \frac{1}{m^2} ds_{AdS_3}^2 + ds^2_8 \right],
\eea
where the warp-factor $\lambda$ is independent of the $AdS_3$ coordinates and $m$ is a constant corresponding to the inverse radius of $AdS_3$. An ansatz for the four-form flux of the form
\be
{F}= {\mathcal I} + m^{-3} \vol(AdS_3) \wedge {\mathcal A},
\ee
will allow us to retain both electric and magnetic fluxes in the decomposition. Here $\mathcal{I}$ is a four-form and $ {\mathcal A}$ denotes a one-form along $M_8$.

After decomposing the gamma matrices, the metric and the fluxes, one may introduce a Killing spinor of the form $\eta = \psi \otimes e^{\lambda/2} \xi$, before utilising the KSE on $AdS_3$
\be
\nabla_{a} \psi = \tfrac{1}{2} \tau_{a} \psi,
\ee
to bring the 11d Killing spinor equation
\be
\nabla_{M} \eta + \frac{1}{288} \left[  \G_{M}^{~NPQR} - 8 \delta_{M}^{~N} \G^{PQR} \right] F_{NPQR} \eta = 0,
\ee
to the form
\bea
\label{KSEeq1}
\left[ \g^{m} \partial_{m} \l + \frac{e^{-3 \l} }{144} \g^{n_1 n_2 n_3 n_4} {\mathcal I}_{n_1 n_2 n_3 n_4} + m \g_9 -\frac{1}{3}\g_9\g^m  e^{-3 \l} {\mathcal A}_m \right] \xi &=& 0, \\
\label{KSEeq2}
\left[ \nabla_{m} - \frac{e^{-3 \l}}{24} \g^{n_1 n_2 n_3} {\mathcal I}_{m n_1 n_2 n_3} - \frac{m}{2} \g_m \g_9 -\frac{1}{4}\g_9 \g_{m}^{~n} e^{-3 \l} {\mathcal A}_n \right] \xi &=& 0.
\eea
Note, in deriving this form, we have used a contraction of (\ref{KSEeq1}) to simplify (\ref{KSEeq2}).

Going further, we consider $M_8$ to be a direct product of $S^2$ and a six-dimensional space $M_6$. The metric and magnetic flux may then be rewritten as
\bea
ds^{2}_{8} &=& e^{2 A} d \Omega_2^2 + ds^{2}_6, \nn
{\mathcal I} &=& \vol(S^2) \wedge \mathcal{H} +  \mathcal {G},
\eea
with $\l$ and $A$ now only depending on the coordinates of the internal six-dimensional manifold $M_6$. The gamma matrices introduced earlier further decompose as
\bea
\g_9 &=& \s_3 \otimes \g_7, \nn
\g_{\a} &=& \s_{\a} \otimes \g_7, \nn
\g_{r} &=& 1 \otimes \g_{r},
\eea
where $\alpha =1,2$ denote the sphere directions and $\g_7 \equiv i \g_{123456}$.

We next consider expanding $\xi$ in terms of a basis of spinors on $S^2$ in a similar fashion to \cite{LLM}
\be
\xi = \chi_+ \otimes \e_+ + \chi_- \otimes \e_-,
\ee
where
\be
\nabla_{\alpha} \chi_{\pm} = \pm i \tfrac{\sigma_{\alpha}}{2} \chi_{\pm}.
\ee
Here $\chi_{+}$ and $\chi_{-}$ are not independent and may be taken to be related by $ \chi_{+} = \sigma_3 \chi_{-}$.

Following the process through, one finds that it is possible to retain the most general flux ansatz at the cost of introducing two spinors $\e_+, \e_-$. In the most general setting, one may imagine these being of indefinite chirality (non-chiral) and linearly independent. In the end one has six equations, divided into two differential relations,
\bea
\label{kse1}
\nabla_{r} \e_{\pm} &=&  \frac{i}{4} e^{-3 \lambda -2 A} {\cal H }_{rs} \g^{s} \e_{\mp} + \frac{e^{-3 \lambda}}{24} {\cal G}_{r stu}  \g^{stu} \e_{\pm} + \frac{m}{2} \g_r \g_7 \e_{\mp} + \frac{e^{-3 \lambda}}{4} \mathcal{A}_{s} \g_7 \g_{r}^{~s} \e_{\mp},\nn
\eea
and four algebraic constraints
\bea
0 &=& \left[ \g^{r} \partial_{r} \l + \frac{e^{-3 \l}}{144}  \g^{rstu} {\cal G}_{rstu} \right] \e_{\pm} + \left[\frac{i}{12} e^{-3 \l -2 A} \g^{rs} {\cal H }_{rs}  + m \g_{7} - \frac{e^{-3 \l}}{3} \g_{7} \g^{r} \mathcal{A}_{r} \right] \e_{\mp}, \nn \label{alg1} \\
0 &=& \left[ \pm \frac{i}{2} \g_7 e^{-A}  + \frac{1}{2} \g^{r} \partial_{r} A \right] \e_{\pm} - \left[ \frac{i}{8} e^{-3 \l-2 A} \g^{rs} {\cal H }_{rs}  + \frac{m}{2} \g_7 - \frac{e^{-3 \l}}{4} \g_7 \g^{r} \mathcal{A}_{r} \right] \e_{\mp}. \nn \label{alg2}
\eea
Note we have multiplied the last algebraic condition by $\g_7$. Adding the algebraic constraints, we can also derive a form independent of ${\cal H }$:
\bea
\label{alg3}
0 &=& \left[ \pm i \g_7 e^{-A} + \g^{r} \partial_{r}(3 \l + A) + \frac{e^{-3 \l}}{48}  \g^{rstu} {\cal G}_{rstu} \right] \e_{\pm} + \left[ 2 m \g_7 - \frac{e^{-3 \l}}{2} \g_7 \g^{r} \mathcal{A}_{r} \right] \e_{\mp}. \nn
\eea
With $\mathcal{A}$ and ${\cal G}$ set to zero, one may use (\ref{alg3}) to derive a projection conditions on the spinors \cite{ LLM,nakwoo}.

Having reduced the KSE, the task remaining is to convert them via spinor bilinears into geometric conditions in order to classify the supersymmetric solutions. In ensuring that the equations of motion are satisfied, we benefit from the well-established observation that the Bianchi identities and the flux equations of motion guarantee the Einstein equations \cite{d11kill}. So, we simply have to impose the Bianchis
\be \label{bianchi} d {\cal H } = d {\cal G} = d \mathcal{A} = 0, \ee and the flux equations of motion\footnote{We have used $\e_{012345678910} = -1$  and Hodge dual
\be
(* \omega)_{\a_1 \cdots \a_q} = \frac{\sqrt{-g}}{p!} \e_{\a_1 \cdots \a_q}^{~~~~~~\b_1 \cdots \b_p} \omega_{\b_1 \cdots \b_p}. \nonumber
\ee
}
\bea
\label{fluxeom}
d (e^{3 \l -2 A} *_6 {\cal H }) &=& - \mathcal{A} \wedge {\cal G}, \nn
d (e^{3 \l +2 A} *_6 {\cal G}) &=& - \mathcal{A} \wedge {\cal H }, \nn
d (e^{3 \l +2 A} *_6 \mathcal{A}) &=&  {\cal H } \wedge {\cal G},
\eea
to claim we have a true supergravity solution.

\section{Torsion conditions}
In this section, we give an account of how the differential and algebraic constraints on the internal $M_6$ may be derived from the KSE, while moving all unsightly expressions to the appendix. We make no assumption about any relationship between $\e_+$ and $\e_-$ and initially regard them as linearly independent and of indefinite chirality. The only simplification arises from the symmetry properties of the inter-twiner $C_6$ defining the charge conjugate of a spinor, $\e^{c} = C_6 \e^{*}$. Here we follow the conventions of \cite{sohnius} and we also adopt $A_6 =1$, so that the six-dimensional gamma matrices satisfy
\bea
\g_m &=& \g_m^{\dagger}, \nn
C^{-1}_{6} \g_m C_6 &=& - \g_{m}^{T},
\eea
where $C_6$ is symmetric and $C_6^{*} = C_6^{-1}$. As a result, $\g_7^{\dagger} = \g_7$ and $\g_7^{T} = - C_6^{-1} \g_7 C_6$. This in turn also dictates which spinor bilinears behave symmetrically under transposition; in general scalars (zero-forms) and three-forms are symmetric, while vectors (one-forms) and two-forms are anti-symmetric.

The general idea in deriving the torsion conditions in  each case is the same. One begins by constructing all spinor bilinears of the form
\be
\bar{\e}_{A} \g^{(p)} \e_{B}, \quad \bar{\e}^{c}_{A} \g^{(p)} \e_{B},
\ee
where the subscripts $A,B$ range for all choices of $+, -$ and $\g^{(p)} \equiv \g^{m_1 \cdots m_p}$. Here $p$ corresponds to the order of the form. In six-dimensions, one need only consider the torsion conditions up to three-forms as the higher forms are simply the Hodge duals of the lower forms. Once one has enumerated all the non-zero forms, one then takes the derivative using (\ref{kse1}) before neatly trying to repackage the derivatives using the algebraic constraints. The results of these calculations can be found in the appendix.

This repetitive and brute force approach to tackling the problem draws on the experience of \cite{jerome1}, where an important constraint on the scalar $f \equiv \bar{\e}^c \e = 0$ arises from demanding consistency between two and three-form torsion conditions. Although we also derive some non-trivial constraints in this manner, as we will see later, most of these conditions may also be found by mining the algebraic constraints (\ref{alg1}), (\ref{alg2}) and (\ref{alg3}).

At each level, we impose consistency amongst the torsion conditions. A flavour for these concerns may be gleaned by commenting on the scalar torsion conditions. Remembering that $\mathcal{A}$ needs to satisfy the Bianchi identify, $d \mathcal{A} = 0$, we see that (\ref{seq2}), (\ref{seq4}) and (\ref{seq9}) are respectively consistent with (\ref{seq11}), (\ref{seq7}) and (\ref{seq3}) under differentiation. Consistency also demands that $X, \Re(Y)$ and $\tilde{Z}$ are constants, a fact that may be read off from other torsion conditions. We also remark that for consistency
\be
d (e^{3 \l} \tilde{K}^4) = d (e^{3 \l} \Re(K^3)) = 0,
\ee
may be expected when one derives the torsion conditions for the vectors at the next level. A quick glance at (\ref{veq1}) and (\ref{veq5}) confirms this to be the case.

However, (\ref{seq5}) is a little unexpected as $\tilde{K}_1$ can be shown to correspond to a Killing direction in the geometry i.e. $\nabla_{(r} \tilde{K}^1_{s)} = 0$. We will comment on this later when we address the Killing directions. For the moment, we stress that the torsion conditions derived in the appendix are completely general and can be adapted to look for supersymmetric solutions with smaller G-structure groups.

Some of the torsion conditions are plainly quite involved and difficult to derive any neat expressions by demanding consistency. However, by differentiating the two-form torsion conditions and comparing with the three-form conditions, one may deduce the following relationships
\bea
\label{cond1} \mathcal{G} W_1 &=& \mathcal{G} \tilde{X}' = 0, \\
\label{cond2} i e^{3 \l - 2 A} \tilde{Z} *_6 {\cal H } &=& \left( 2m e^{3 \l +A}  Z+ i e^{3 \l} W_2 \right) {\cal G}, \\
\label{cond3} \tilde{X} e^{3 \l} {\cal G} &=& e^{3 \l -2 A} \Re(Y) *_6 {\cal H }.
\eea
Interestingly, differentiating (\ref{cond2}), one can use the flux equations of motion (\ref{fluxeom}) and the scalar torsion conditions to show that the resulting expression is satisfied. This we take as a heart-warming sign of consistency.  Whereas the derivative of (\ref{cond3}) leads to
\be
\Re(K^3) \wedge \mathcal{G} = 0,
\ee
but it is difficult to comment on the significance of this expression.

\subsection{Algebraic constraints}
Next we turn to the algebraic constraints. Similar conditions on the scalar bilinears may be derived from (\ref{alg1}), (\ref{alg2}) and (\ref{alg3}). In general, when all fluxes are non-zero, we find the following conditions
\bea
\label{genconst}
\tilde{Z} &=& W_1 = \tilde{X}' = \Re(Y) = d X = 0, \nn
ie^{-A} W_2 &=& - 2 m Z, \nn
e^{-A} \tilde{X} &=& -2 m \Im(\tilde{Y}), \nn
2m X + e^{-A} \Im(Y) &=& \frac{e^{3\l}}{2} \Im(L^3) \lrcorner *_6 {\cal G} + \frac{e^{-3 \l}}{2} \mathcal{A} \lrcorner K^1,
\eea
where the last expression simplifies when $\mathcal{A} = \mathcal{G} = 0$. Combining these expressions with the conditions derived from the torsion conditions, we note that the algebraic constraints imply the constraints derived from the torsion conditions, up to the existence of the expression
\be
\label{cond4}
\tilde{X} \mathcal{G} = 0.
\ee
Interestingly, this says that $\tilde{X}$ necessarily has to be zero if we are to support a magnetic four-form flux purely on $M_6$. It is not possible to derive this expression from the algebraic constraints directly.

\subsection{Killing directions}
From direct application of (\ref{kse1}), one notes that there are three candidate Killing directions satisfying the Killing equation $\nabla_{(r} K_{s)} = 0$: $\tilde{K}^{1}, \Im(\tilde{K}^{3})$,  and $K^4$ all satisfy this condition. In this subsection, we ask whether the three of these generate also symmetries of the two warp factors $\lambda, A$ and the three flux terms $\mathcal{A}, \mathcal{H}$ and $\mathcal{G}$.

We begin with the warp factors. Using (\ref{alg1}) it can be shown that the overall warp factor $\lambda$ is independent of these directions\footnote{Here $\mathcal{L}_{X}$ denotes the Lie derivative w.r.t. a vector $X$, $\mathcal{L}_{X} \equiv d i_{X} + i_{X} d$.}:
\be
{\cal L}_{K} \lambda \equiv i_{K} d \lambda = 0,~~~~ K \in \{\tilde{K}^{1}, \Im(\tilde{K}^{3}), K^4 \}.
\ee
However, from (\ref{alg2}), one finds that the size of the $S^2$ depends on $\tilde{K}^{1}$:
\bea
\label{kill1} (\tilde{K}^1)^{r} \partial_{r} A &=& - e^{-A} X', \\
 (K^{4})^{r} \partial_{r}A &=& 0, \\
\Im(\tilde{K}^3)^r \partial_{r} A &=& 0. \eea

We now proceed onto the flux terms. It is possible to show
\bea
{\cal L}_{K} \mathcal{A} &\equiv& d i_{K} \mathcal{A} = 0,\nn
{\cal L}_{K} \mathcal{G} &\equiv& d i_{K} \mathcal{G} = 0
~~~~ K \in \{\tilde{K}^{1}, \Im(\tilde{K}^{3}), K^4 \}.
\eea
The upper line here follows from (\ref{alg1}), while (\ref{2eq3}), (\ref{2eq7}) and (\ref{2eq11}) are used in deriving the result in the lower line. In both cases we also make use of the Bianchis (\ref{bianchi}).

Finally, it can be shown that $d i_{K^{4}} {\cal H } = d i_{\Im(\tilde{K}^3)} {\cal H } = 0$ by using (\ref{alg2}) to derive
\bea
 i_{\Im(\tilde{K}^{3})} {\cal H } &=& \Re(Y) e^{2A} \mathcal{A} - 2m e^{3 \l + 2 A} \Re(K^3) - 2 e^{3 \l + 2 A} \tilde{X} dA, \nn
i_{K^{4}} {\cal H } &=& -2 i e^{3 \l+2A} \left(m \tilde{K}^{4} - W_2 d A \right),
\eea
which can be shown to close using the torsion conditions. In a similar fashion, it can be shown that $\mathcal{L}_{\tilde{K}^1} \mathcal{H} \neq 0$:
\bea
\label{kill2}
d \left( i_{\tilde{K}_{1}} {\cal H } \right) &=& -2 d(e^{3 \l +A} \Im(K_{3})) = -2 e^{-A} X'{\cal H }.
 \eea
  Therefore, from the three candidate Killing directions, we reach the conclusion that two of them will generate symmetries of the 11d solution i.e. they are symmetries of the metric, warp factors and fluxes. In the case of $\tilde{K}^1$, (\ref{kill1}) and (\ref{kill2}) seem to be innocuous and one imagines one can promote $\tilde{K}^1$ to a full isometry by setting $X'= 0$. However, this leads to other problems. Via (\ref{seq5}), setting $X'=0$ also sets $\tilde{K}^1=0$.

\subsection{Summary of constraints}
Before proceeding to the next section, here we collect and enumerate the constraints that have arisen as a result of either the torsion conditions, the algebraic constraints or from demanding consitency. In effect, any supersymmetric, well-defined geometry should satisfy the following:
\bea
\label{constsummary}
\tilde{Z} &=& W_1 = \tilde{X}' = \Re(Y) = d X = 0, \nn
ie^{-A} W_2 &=& - 2 m Z, \nn
e^{-A} \tilde{X} &=& -2 m \Im(\tilde{Y}), \nn
2m X + e^{-A} \Im(Y) &=& \frac{e^{3\l}}{2} \Im(L^3) \lrcorner *_6 {\cal G} + \frac{e^{-3 \l}}{2} \mathcal{A} \lrcorner K^1, \nn
\tilde{X} &=& 0 \quad \mbox{OR} \quad \mathcal{G} = 0, \nn
X' &=& \tilde{K}^{1} = 0.
\eea

\textbf{Aside}

Although the constraints of the last line of this summary seem unnecessarily strong, we can however try to relax this constraint to consider non-zero $X'$ and $\tilde{K}^1$. Combining (\ref{kill1}) and (\ref{seq5}), we find two relations
\bea
 d_{5} \left( e^{-A} X' \right) &=& 0, \nn
\alpha \partial_{\psi} X' &=& e^{-A} \left[ 1 - (X')^2 \right], \label{eqnabove}
\eea
where we have used $d \equiv d_5 + \tilde{K}_1 (\tilde{K}^{1r} \partial_{r}) $ and have defined a dual vector $\tilde{K}^1 = \alpha \partial_{\psi}$ with $\alpha$ constant. Here the direction $\psi$ corresponds to a Killing direction on the transverse space $M_6$. By using (\ref{kill1}) it is possible to write $X'$ as
\be
X' = - {\alpha}{\partial_{\psi} e^{A}},
\ee
so that the above two equations (\ref{eqnabove}) may be rewritten as
\bea
d_5 \partial_{\psi} A &=& 0, \\
\alpha^2 \partial_{\psi}^{2} e^{A} &=& e^{-A} \left[\alpha^{2} (\partial_{\psi}e^{A})^2  - 1\right]
\eea
Now, we see that either $d_5 A = 0$ or $ \partial_{\psi} A = 0$. In the latter case, one is forced to adopt $X' = \tilde{K}^1 = 0$, while in the former case, the warp factor $A$ depends only on $\psi$ and is independent of the other directions. Indeed, there is a simple solution and one can show
\be
e^{A} \sim \sin (\psi).
\ee
So, in general, the radius of the $S^2$ depends on an internal Killing direction $\psi$. Though it would be interesting to follow this case further, for simplicity, we decide to set  both $X'$ and $\tilde{K}^1$ to zero. The condition $X' = 0$ does not seem much of a sacrifice, as it simply states that the norms of $\e_+$ and $\e_-$ should be the same. Combining them with (\ref{seq3}), we also see that the spinors have constant norm. Henceforth, it is to be understood that we have specialised to the case where the spinors have constant norm.

\section{Killing spinors with $SU(2)$-structure}
Before looking at solutions with two orthogonal chiral spinors defining $SU(2)$-structure, as a suitable warm-up exercise, we may briefly look at the case where $\e_+$ and $\e_-$ can be expressed in terms of a single unit norm, chiral spinor $\eta$. In general, it is difficult to see this directly from the KSE, however with the torsion conditions already derived, the result becomes immediate.

We start by adopting the following projection conditions for $\eta$
\bea
\g_{12} \eta &=& \g_{34} \eta = \g_{56} \eta = -i \eta, \quad \Rightarrow \g_7 \eta = -\eta, \nn
\g_{135} \eta &=& \eta^c,
\eea
so that the forms defining the $SU(3)$-structure $J$ and  $\Omega$ are given by
\bea
J &=& \frac{i}{2} \bar{\eta} \g_{mn} \eta ~e^{mn} = e^{12} + e^{34} + e^{56}, \nn
\Omega &=& \frac{1}{3!} \bar{\eta}^c \g_{mnp} \eta ~e^{mnp} = (e^1 + i e^2)(e^3+ie^4)(e^5+ie^6), .
\eea
where we have omitted the customary wedge products for brevity. As one chiral spinor defines a single real scalar, we can choose $\e_{+} = f \eta$, where $f$ is a real function. For $\e_-$, we choose the linear combination
\be
\e_- = g_1 \eta + g_2 \eta^c,
\ee
where in general, we allow $g_i$ to be complex functions. The scalar constraints derived previously enforce the choice
\be
\e_{+} = \eta, \quad \e_- = - i \eta,
\ee
where we have taken $X=1, X'= 0$ and the relative sign above is necessitated so that \be \label{su3cond} e^{-A} = 2 m.\ee As the vectors disappear, the other warp factor $\lambda$ can also be seen to be constant due to the scalar torsion conditions. One may then set it to zero. Also since $X$ is non-zero, we also have $\mathcal{A} = 0$ from (\ref{seq9}).

The absence of any vectors means that the RHS of the vector torsion conditions become algebraic leading to a considerable simplification. Then, we observe that $ \mathcal{G} = 0$ from (\ref{veq2}) and (\ref{veq3}), (\ref{veq7}) and (\ref{su3cond}) lead to the same expression for $\mathcal{H}$
\be
\mathcal{H} = \frac{1}{2 m} J.
\ee
The remaining torsion conditions can be shown to be satisfied if the $M_6$ is Calabi-Yau
\be
d J = d \Omega = 0.
\ee
The reader may observe that when $m=1$ corresponding to unit $AdS$ radius, the relative factor between the $AdS_3$ and $S^2$ factors agrees with \cite{MSW}.

Having warmed up, going beyond $SU(3)$-structure, one may ask are there any solutions with two orthogonal, chiral spinors $\eta_1, \eta_2$:
\be \bar{\eta}_{i} \eta_j = \delta_{ij}, \quad \g_7 \eta_i = - \eta_i. \ee
Together these two spinors define a canonical $SU(2)$-structure in 6d as explained in \cite{jerome1,jerome5}. Such a structure may equivalently be specified by two one-forms $P^1$, $P^2$ and three two-forms $J^m$ given in terms of $\eta_i$:
\bea
J^{m} &=& - \frac{i}{2} \sigma_{m}^{ij} \bar{\eta}_{i} \g_{(2)} \eta_{j}, \nn
P^{1} - i P^2 &=& - \frac{1}{2} \e^{ij} \bar{\eta}^{c}_{i} \g_{(1)} \eta_{j},
\eea
where $\s_{m}$ denote the Pauli matrices. We may also define
\bea
\Omega \equiv J^2 + i J^1, \nn
J \equiv J^3,
\eea
and specify the $SU(2)$-structure in terms of ($J, \Omega, P^1,P^2$). The 6d metric then takes the form:
\be
ds^2 = e^i e^i + (P^1)^2 + (P^2)^2.
\ee

Having introduced $\eta_1, \eta_2$ and the $SU(2)$-structure, the task now remains to connect these to the original spinors $\e_+, \e_-$ appearing in the KSE. Following \cite{jerome1}, we adopt the same expression for a general non-chiral spinor $\e_+$, while making no assumptions about $\e_-$:
\bea
\label{2spinors}
\e_+ &=& \cos \theta \eta_1 + \sin \theta \left(c \eta_1^c + d \eta_2^c \right), \nn
\e_- &=& a_1 \e_+ + a_2 \g_7 \e_+ + a_3 \e_+^c + a_4 \g_7 \e_{+}^c,
\eea
Here, as explained in \cite{jerome1}, $c, d$ are complex functions satisfying $|c|^2+|d|^2 = 1$, while $a_i$ denote arbitrary complex functions that we have introduced. Note also that $\e_{+}$ is of indefinite chirality as \be \g_7 \eta_i = - \eta_i ~~ \Rightarrow ~~\g_7 \eta_i^c = \eta_i^c. \ee We also note that $\e_+$ and $\e_-$ are now linearly dependent, as one can always write
\be
\e_- = A \e_+ + B \g_7 \e_+ + C \e_+^c + D \g_7 \e_+^c.
\ee

Our task in this section is to determine the most general way in which the functions $a_i$ in (\ref{2spinors}) may be chosen consistent with the constraints derived in the previous section and the $SU(2)$-structure. We begin by recalling from (\ref{cond4}) that there are two cases to be considered: either $\tilde{X} = 0$ or $ \mathcal{G} = 0$. We now proceed case by case. \\

\textbf{Case 1: } $\tilde{X} = 0, ~ \mathcal{G} \neq 0 $ \\

Following our earlier choice of $\e_+$, with $SU(2)$-structure, we may now define three scalars
\be
\bar{\e}_{+} \e_+ = 1, \quad \bar{\e}_{+} \g_7 \e_+ = - \sin \zeta, \quad \bar{\e}^{c}_{+} \e_+ = c \cos \zeta,
\ee
where we have employed the following relabelling for later convenience
\be
\sin \zeta = \cos (2 \theta), \quad \cos \zeta = \sin (2 \theta).
\ee
We may also define two real vectors
\be
V^m_1 = \bar{\e}_+ \g^{m} \e_+, \quad V_2^{m} = i \bar{\e}_+ \g^{m} \g_7 \e_+.
\ee
In terms of these scalars and vectors, we have reproduced the scalar constraints and the vectors in the appendix. We start by noting that $\tilde{X} \equiv \bar{\e}_{+} \g_7 \e_+ = 0$ implies that $\sin \zeta = 0$, and from (\ref{seq4}) and $\Re(Y) = 0$ we immediately see that $\Re(K^3) = 0$, telling us that
$ \Re(a_1) = \Im(a_2) = 0$. Then from the condition that $\tilde{K}^1 = 0$ (\ref{scalarcond2}), we infer that
\be
a_{4} = - c a_2 = - c a, \quad -a_3 c^* = a_1 = i b, \mbox{ where } a,b \in \mathbb{R}.
\ee
Then demanding consistency between the $W_1 = 0$ and $\bar{\e}_- \e_- = 1$ expressions in (\ref{scalarcond1}) leads to
\be
a = b = 0 \mbox{ OR } |c|^2 = 1 \Rightarrow SU(3)\mbox{-structure},
\ee
thus ruling out this possibility. Therefore, we can rule out the possibility of a magnetic flux $\mathcal{G}$ in these backgrounds with $SU(2)$-structure. We expect that this statement may be generalisable to all M-theory geometries with $AdS_3 \times S^2$ factors, but it is difficult to extend this argument without referring to specific cases. Recall also that this is a stronger statement than that of \cite{LLM}, where the presence of a similar flux term was ruled out only perturbatively using the KSE. It would be interesting to analyse the torsion conditions corresponding to LLM. \\

\textbf{Case 2: }  $\mathcal{G} = 0, ~ \tilde{X} \neq 0$ \\

Though it would be desirable to determine the most general form of the spinor, we find that this is a formidable task. One would need to assume that the $a_i, ~i=1,...,4$ are general complex functions and then hope to use the torsion conditions as derived to reach some conclusion as to which $a_i$ are non-zero (see the appendix \cite{Gauntlett:2005ww} for a simpler set-up with two functions.). So to make the problem more tractable, in the appendix we make the assumption that the Killing direction aligns itself with only one of the vectors defining $SU(2)$-structure. With this assumption, we find that the scalar constraints allow for a general relationship of the form
\be
\label{genspinor}
\e_{-} = i \frac{e^{-A}}{2m} ( \sin \zeta \g_7 \e_{+} - c \cos \zeta \e_+^c).
\ee

Also, as the norm of $\e_-$ is unity $\bar{\e}_- \e_- = 1$ ($X=1, ~X'=0$), this means that
\be
\label{genspinrel}
(2 m e^{A})^2 = \sin^2 \zeta + |c|^2 \cos^2 \zeta.
\ee

The next step in the analysis is to plug this spinor back into the torsion conditions to see if it survives the integration process. Before doing this, we  can immediately ask what fluxes may be supported by this general spinor condition. One notes that as $K^1 = \Re(\tilde{Y}) = 0$, but $X=1$, then (\ref{seq9}) tells us that $\mathcal{A} = 0$. We then move onto (\ref{veq2}) which again with $\Re(\tilde{Y}) =0$, confirms that the presence of $\mathcal{G}$ is intrinsically linked to $\Im(L_3)$. A quick calculation with the above spinor then confirms that $\Im(L_3) = 0$ and, as expected, $\mathcal{G} = 0$. So,  in the general class of solutions corresponding to (\ref{genspinor}), we observe that the only non-zero flux term is $\mathcal{H}$.

Also, despite not being able to completely rule out the presence of an electric flux term $\mathcal{A}$, given our assumptions about Killing directions, we have found that it is not possible to switch it on. We note that the case we have treated here almost falls into the K\"{a}hler-4 of \cite{eoin}. Indeed, one can compare (2.5) of that paper directly with (\ref{seq9}), where the $d \rho$ direction corresponds to $K^1$. However, as we have imposed $K^1 = 0$, we see that our class of solutions cannot fall into the K\"{a}hler-4 class.

We also remark that when $c$ is non-zero, in contrast to the work in the next section, one notices that the two-from bilinears will all mix the $SU(2)$-structure two-forms $J$ and $\Omega$. This will lead to complicated conditions on the geometry arising from (\ref{2eq3}), (\ref{2eq7}) and (\ref{2eq11}) even before one starts testing the other two and three-form torsion conditions. It is extremely unlikely that $c \neq 0$ will lead to an integrable solution, so we decide not to pursue this avenue. Interestingly, if one relaxes the G-structure, then it may be possible to integrate these equations. However, we are acutely aware that there is no guarantee of finding explicit solutions \cite{Gauntlett:2005ww}.

Thus, setting $c = 0$ so that the $SU(2)$-structure forms $J, \Omega$ do not mix, the Killing direction is just along $\Im(\tilde{K}^3)$, and one recovers, as we will verify, the work of \cite{wrappedbranes,nakwoo}. Note, the supersymmetry conditions for this class of solutions were originally derived in \cite{wrappedbranes}.

One final remark before we move on. From (\ref{genspinrel}), with $c=0$, we
\be
\label{sinz}
2 m e^{A} = \pm \sin \zeta,
\ee
and the relationship between $\e_+$ and $\e_-$ simply reads \be \label{finalspinor} \e_- = \pm i \g_7 \e_+. \ee A similar relationship also featured in \cite{LLM}. It may be arrived at by noting that the Killing spinor equations when $\mathcal{A} = \mathcal{G} = 0$, may be rewritten in the form:
\bea
\nabla_{r} \tilde{\e} &=&  \frac{a}{4} e^{-3 \lambda -2 A} H_{rs} \g^{s} \g_7 \tilde{\e} - \frac{iam}{2} \g_r \tilde{\e}, \nn
0 &=&  \g^{r} \partial_{r} \l \tilde{\e} + \frac{a}{12} e^{-3 \l -2 A} \g^{rs} \g_7 H_{rs} \tilde{\e} - ia m  \tilde{\e},  \nn
0 &=&  \frac{i}{2} \g_7 e^{-A} \tilde{\e}  + \frac{1}{2} \g^{r} \partial_{r} A \tilde{\e} - \frac{a}{8} e^{-3 \l-2 A} \g^{rs} \g_7 H_{rs}  \tilde{\e} + \frac{i am}{2} \tilde{\e}, \nn
0 &=& i \g_7 e^{-A} \tilde{\e} + \g^{r} \partial_{r}(3 \l + A) \tilde{\e}  - 2 iam \tilde{\e},
\eea
where \be \tilde{\e} = \e_+ + a i \g_7 \e_{-}, \quad  a = \pm 1. \ee

We stress that from the offset we have made no assumption about the nature of the fluxes. We have arrived at the above conclusions by adopting various simplifying assumptions which we have attempted to motivate at each stage. Here we recap our assumptions. We began by imposing $SU(2)$-structure on the internal space $M_6$. Our supersymmetry analysis suggests that this is enough to kill the possibility of the flux $\mathcal{G}$. The assumptions we have made about the Killing directions; namely 1) $X', \tilde{K}^1 = 0 \Rightarrow K^1 = 0$ and 2) the Killing direction aligns itself with either $V_1$ or $V_2$ are strong enough to determine the relationship between $\e_+$ and $\e_-$ uniquely, while at the same time, setting $\mathcal{A} = 0$. Finally, as we are unaware of any integrable class of solutions with $SU(2)$-structure and $c \neq 0$, we have set $c=0$ leading to an LLM type relationship between the two linearly dependent spinors. In the next section, we derive the constraints on the geometry for this class of solutions. We recover the work of \cite{wrappedbranes,nakwoo} before making some comments on the properties of this class.

\section{Geometries with $SU(2)$-structure}
In the previous section, we have provided support for (\ref{finalspinor}) being the only relationship between $\e_-$ and $\e_+$ so that $M_6$ admits $SU(2)$-structure. This relationship between the spinors was initially assumed in \cite{nakwoo} as a starting point for a G-structure treatment using a simplified flux ansatz. In contrast to \cite{nakwoo}, we have taken a step backwards by relaxing some constraints, but have yet been led to the the same relationship. In this section, we analyse this case in detail from a slightly different perspective by imposing $SU(2)$-structure and using the torsion conditions. One upshot of our analysis is that we will be able to dispense with the messy Fierz identities that featured in \cite{nakwoo} and will see how the torsion conditions specify the conditions on the geometry through the G-structure group. The agreement with the results of \cite{nakwoo} provide a highly non-trivial check of their results, and in turn, of the results of \cite{wrappedbranes} where the class of solutions originally appeared.

We now proceed to derive the conditions on this class by placing the spinor relation $\e_{+} = - a i \g_7 \e_-$ into our torsion conditions. We adopt the sign choice $a=1$ so that $\e_- = i \g_7 \e_+$. With this choice the non-zero bilinears become:
\bea
W_1 &=& \bar{\e}^{c} \e, \quad
X = \Im(\tilde{Y}) = \bar{\e} \e, \quad
\tilde{X} = \Im(Y) = \bar{\e} \g_7 \e, \nn
K^2_m &=& \Im(\tilde{K}^3)_m = \bar{\e} \g_m \e, \quad
\tilde{K}^2_m = K^3_m = i \bar{\e} \g_{m} \g_7 \e, \nn
L^1_{rm} &=& \tilde{L}^3_{rm} = i \bar{\e} \g_{rm} \e, \quad
\tilde{L}^1_{rm} = L^3_{rm} = i \bar{\e} \g_{rm} \g_7 \e, \quad
L^4_{rm} = i L^5_{rm} = i \bar{\e}^{c} \g_{rm} \g_7 \e, \nn
M^2_{rmn} &=& -i (*_6 M^3)_{rmn} = i \bar{\e} \g_{rmn} \e, \quad
M^4_{rmn} = (*_6M^6)_{rmn} = i\bar{\e}^{c} \g_{rmn} \e,
\eea
where we have dropped the redundant subscript on the spinor.

The scalar torsion conditions simplify to
\bea
\label{0tor}
d X &=& 0, \nn
e^{-3 \l} d (e^{3 \l} \tilde{X} ) &=& 2 m \Re(K^3), \nn
d(e^{-A} \tilde{X}) &=& 0, \nn
e^{-3 \l} d (e^{3 \l + A} X) &=& -\Re(K^3).
\eea

Now, as $c=0$, then $d=1$ and using $X=1$, $\e$ takes the form
\be
\label{nspinor}
\e = \cos \theta \eta_1 + \sin \theta \eta_2^c.
\ee
Note, the two contributions to the spinor still have opposite chirality, so $\e$ is still of indefinite chirality. As in \cite{jerome1}, calculations may be simplified by the introduction of set projection conditions:
\bea
\g_{12} \eta_1 = \g_{34} \eta_1 = - \g_{56} \eta_1 &=& i \eta_1, \quad \g_{135} \eta_1 = - \eta^{c}_1, \nn
-\g_{12} \eta_2 = -\g_{34} \eta_2 = - \g_{56} \eta_2 &=& i \eta_2, \quad \g_{135} \eta_2 = - \eta^{c}_2,
\eea
where the two spinors are related via
\be \g_5 \eta^{c}_{2} = \eta_1. \ee
With this choice, the structure forms become $J^1 = e^{14} +e^{23}, J^2 = e^{13}-e^{24}, J^3 = e^{12} + e^{34}, P^1 = e^5$ and $P^2 = e^6$.

Then inserting the above spinor (\ref{nspinor}) into the bilinears appearing in the scalar torsion conditions, one establishes that:
\bea
\Re(K^3) &=& -P^2 \cos \zeta, \nn
\tilde{X} &=& -\sin \zeta.
\eea
Then one observes that the torsion conditions (\ref{0tor}) may be integrated neatly to give:
\bea
\label{intscal}
2m y &=& e^{3 \l} \sin \zeta, \nn
P^{2} &=& e^{-3 \l} \sec \zeta d y, \nn
e^{A} &=&  \frac{\sin \zeta}{2 m}.
\eea
Note, the last expression here is consistent with (\ref{sinz}). The above also mean that $y = e^{3 \l +A}$.

We may now turn attention to $\Im(\tilde{K}^{3})$. As has been shown previously, it is a Killing direction and generates a symmetry of the entire supergravity solution. In addition, $\mathcal{L}_{\Im(\tilde{K}^{3})} P^2 = 0$ may be shown from (\ref{0f3}) and by use of the 6d Fierz identity as was illustrated in \cite{Donos1,nakwoo}.

Since $\Im(\tilde{K}^3) = P^1 \cos \zeta$ is Killing, we can introduce local coordinates via a vector $\Im(\tilde{K}^3) = C^{-1} \partial_{\psi}$ such that
\be
P^1 = C\cos \zeta (d \psi + \rho),
\ee
with $\rho = \rho_{i}(x^j,y)dx^i$ and $C$ an arbitrary constant. The 6d metric may then be expressed as
\be
ds^2 = g^{4}_{ij} dx^{i} dx^{j} + e^{-6 \l } \sec^2 \zeta dy^2 + C^2 \cos^2 \zeta (d \psi + \rho)^2,
\ee
where $g^4_{ij}, \l$ and $\zeta$ are, in principle, functions of $x_i$ and $y$.

Next one can move onto the one-form torsion conditions. In this simplified setting they become:
\bea
d(e^{3 \l +A} K^2) &=& - e^{-A} \Im(Y) \mathcal{H} + e^{3 \l} \tilde{L}^1, \label{0f1} \\
\label{0f2} d(e^{6 \l +A} \tilde{K}^2) &=& 0, \quad d(e^{3 \l} \Re(K^3)) = 0, \\
d (e^{6 \l + 2A} \Im(\tilde{K}^3)) &=& - e^{3 \l} \tilde{X} \mathcal{H} + 2 m e^{6 \l + 2 A} L^1 + 2 e^{6 \l +A} \Re(L^3). \label{0f3}
\eea
The two equations appearing in  (\ref{0f2}) are not independent and are trivially satisfied through $y =  e^{3 \l +A}$. It is also possible to show that (\ref{0f3}) is trivially satisfied when (\ref{0f1}) is inserted, making them also consistent. As a result, $\mathcal{H}$ may be determined from (\ref{0f1}):
\bea
\mathcal{H} &=& \frac{1}{2m} \left[ e^{3 \l} (P^1 \wedge P^2 - J \sin \zeta) + C d \left(e^{3 \l +A} \cos^2 \zeta  (d \psi + \rho) \right)\right], \nn
&=& - y J - \frac{C}{2m} \partial_{y} (y \sin^2 \zeta) d y \wedge (d \psi + \rho) - \frac{Cy}{m} \cos \zeta \sin \zeta d_4 \zeta \wedge (d \psi + \rho) \nn
&+& \frac{C y \cos^2 \zeta}{2m} d \rho. \label{Hflux}
\eea
As can be readily seen from, for example  (\ref{2eq3}), $\mathcal{H}$ is closed and thus satisfies the Bianchi identity.

In terms of the $SU(2)$-structure, we may rewrite the remaining torsion forms as
\bea
L^1 &=& -J + P^1 \wedge P^2 \sin \zeta, \nn
\tilde{L}^1 &=& J \sin \zeta - P^1 \wedge P^2 , \nn
L^4 &=& -i \Omega \cos \zeta, \nn
M^2 &=& -J \wedge P^1 \cos \zeta, \nn
M^3 &=& - i J \wedge P^2 \cos \zeta, \nn
M^4 &=& i \Omega \wedge \left(P^1-i P^2 \sin \zeta \right), \nn
M^6 &=& \Omega \wedge \left( -P^1 \sin \zeta + i P^2\right). \label{remaintor}
\eea

The non-zero two-form torsion conditions become:
\bea
d (e^{6 \l + 2 A} L^1) &=& e^{3 \l} \Re(K^3) \wedge \mathcal{H} - 2 e^{6 \l +A} *_6 M^2, \label{2f1} \\
d (e^{3 \l} \tilde{L}^1) &=& 0,  \label{2f2} \\
d (e^{6 \l +A} \Re(L^3)) &=& 2m e^{6 \l +A} *_6 M^2, \label{2f3}\\
d(e^{3 \l +A} \Re(\tilde{L}^3)) &=& e^{-A} \tilde{K}^2 \wedge \mathcal{H} + e^{3 \l} \Im(M^3), \label{2f4}\\
d(e^{6 \l +A} L^5) &=& 2m e^{6 \l +A} M^4 + i e^{6\l} M^6, \label{2f5}\\
d(e^{3 \l} L^4) &=& 0. \label{2f6}
\eea
Thus, we have from (\ref{2f2}) and (\ref{2f6}) that
\bea
d (e^{3 \l} \cos \zeta \Omega) &=& 0, \label{Om_diff} \\
2m d (e^{3 \l +A} J) - C d_{4} \rho \wedge dy &=& 0 \label{J_diff},
\eea
where $d \equiv d_4 + d y \wedge \partial_y + d \psi \wedge \partial_{\psi}$. We may now rescale the base metric $g^{4}_{ij} \rightarrow  e^{-3 \l -A} \tilde{g}^{4}_{ij}$ such that
\bea
\label{rescaledOm_diff} d (\cot \zeta \tilde{\Omega}) &=& 0, \\
\label{rescaledJ_diff} \partial_{y} \tilde{J} &=& \frac{C}{2m} d_4 {\rho}, \quad d_4 \tilde{J} = \partial_{\psi} \tilde{J} = 0.
\eea
It is possible to show that the other two-form torsion conditions above are satisfied using (\ref{Hflux}) and (\ref{remaintor}).

Following along in the steps of \cite{jerome1}, we note that the rescaled base  $\tilde{M}_4$ is a complex manifold and with $d_4 \tilde{J} = 0$, we have locally a family of K\"{a}hler metrics on $\tilde{M}_4$ which are parametrised by $y$. Given a family of such metrics, we may also write
\be
d_4 \tilde{\Omega} = i \tilde{P} \wedge \tilde{\Omega},
\ee
where $\tilde{P}$ is the canonical Ricci-form connection. In other words
\be
\tilde{R} = d_4 \tilde{P},
\ee
where $\tilde{R}_{ij} = \tfrac{1}{2} \tilde{R}_{ijkl} \tilde{J}^{kl}$ denotes the Ricci-form.
From (\ref{Om_diff}), we can now infer that
\be
(\tilde{P} + i d_4 \log \tan \zeta ) \wedge \tilde{\Omega} = 0 \Rightarrow \tilde{P}= - \tilde{J} \cdot d_{4} \log \tan \zeta,
\ee
or that $\tilde{P} + i d_4 \log \tan \zeta$ is a $(1,0)$ form on $M_4$. Using $\tilde{\Omega} \wedge \bar{\tilde{\Omega}} = 4 \tilde{vol}_4$, we further deduce that
\be
\partial_{y} \log \sqrt{\tilde{g}} = 2 \partial_{y} \log \tan \zeta.
\ee
Then citing an identity valid for self-dual two forms $\omega^{+}$ on $M_4$ when the complex structure $\tilde{J}$ is independent of $y$,
\be
(\partial_{y} \tilde{J})^{+} = \frac{1}{2} \partial_y \log \sqrt{\tilde{g}} \tilde{J},
\ee
we arrive at
\be
(d_4 \rho)^{+} = \frac{2m}{C} \partial_{y} \log \tan \zeta \tilde{J}.
\ee
Finally, we enumerate the three-form torsion conditions:
\bea
d(e^{6 \l +A} *_6 M^2) &=& 0, \label{3f1} \\
\label{3f2} d(e^{6 \l} M^6) &=& 2 m i e^{6 \l} *_6 L^4, \\
\label{3f3} d(e^{6 \l+A} M^4) &=& e^{6 \l} *_6 L^4.
\eea
Here (\ref{3f1}) follows from (\ref{2f1}) or (\ref{2f3}), while (\ref{3f2}) and (\ref{3f3}) tell us that
\bea
\tilde{\Omega} \wedge d_4 \rho &=& 0, \nn
m C y \cos^2 \zeta \partial_{y} \rho &=&  \tilde{J} \cdot d_4 \log \cos \zeta \label{delyrho}.
\eea
Combining this with earlier results, we then have
\be
\tilde{P} = m C y \cot^2 \zeta \partial_{y} \rho .
\ee
This exhausts the complete set of torsion conditions that we have derived. Now, provided the Killing spinor is \textit{not} null\footnote{Recall that a Killing spinor $\e$ is null if the vector bilinear $V^{\mu}$ constructed from $\e$, $V^{\mu} = \bar{\e} \g^{\mu} \e$,  has zero-norm i.e. $V_{\mu} V^{\mu} = 0$. }, supersymmetry will guarantee the Einstein equations if we ensure that the Bianchi equation and the flux equation of motion, $d ( e^{3 \l - 2 A} *_6 H) = 0$, are satisfied \cite{d11kill}. The fact that the Killing spinor is not null may be deduced from the Fierz identity as in \cite{Donos1,nakwoo} where it is shown that although the combination $P^1 + P^2$ is null, both $P^1$ and $P^2$ have non-zero norms. From (\ref{Hflux}) we remember that the Bianchi is satisfied by construction. A calculation reveals the flux equation to be also satisfied through the use of (\ref{Om_diff}) in the form
\be
d \left( y^2 \frac{\cos^2 \zeta }{\sin^2 \zeta} J \wedge J \right) = 0.
\ee

\subsection{Some remarks}
To make contact with the work appearing in \cite{nakwoo} we simply need to compare the above rescaled metric
\bea
\label{metric1}
ds^2 &=& e^{2 \l} \biggl[ \frac{1}{m^2} ds^2(AdS_3) + e^{2 A} ds^2(S^2) + e^{-3 \l -A}\tilde{g}^{4}_{ij} dx^{i} dx^{j} \nn &+& e^{-6 \l } \sec^2 \zeta dy^2 + C^2 \cos^2 \zeta (d \psi + \rho)^2 \biggr],
\eea
with the analytically continued metric (3.6) of \cite{nakwoo}:
\bea
\label{metric2}
ds^2 &=& e^{2 \tilde{A}} ds^2(S^2)  + \tilde{y}^2 e^{-\tilde{A}}ds^2(AdS_3) + e^{-\tilde{A}} h_{ij} dx^{i} dx^{j} \nn
&+& \frac{\tilde{y}^2 e^{-\tilde{A}} }{4} \cos^2 {\xi} (d \psi +V) + \frac{4 e^{2\tilde{A}} }{\tilde{y}^2 \cos^2 {\xi}} d\tilde{y}^2,
\eea
where \be \tilde{y} = e^{B + \tilde{A}/2} = \frac{2 e^{3 A/2}}{\sin {\xi}} .\ee The above metric is the result of analytic continuation from the original metric (2.1), (2.40) of \cite{nakwoo} with the following identifications:
\bea
\psi &=& t, \quad \tilde{y}^2 \cos^2 {\xi} = -4 e^{3A} \cosh^2 \zeta, \nn
\tilde{y} &=& i y, \quad e^{B} = i e^{B}, \quad \tilde{A} = A.
\eea
The above metrics (\ref{metric1}) and (\ref{metric2}) may be identified via:
\bea
\l &=& B, \quad A = \tilde{A} - B, \quad y = \tilde{y}^2, \nn
\zeta &=& {\xi}, \quad \rho = V, \quad C = \frac{1}{2}, \quad m = 1.
\eea

One may also identify (\ref{rescaledOm_diff}), (\ref{rescaledJ_diff}) and (\ref{delyrho}) with (2.83), (2.85) and (2.86) of \cite{nakwoo} after one takes account of the change of variables inherent in the analytic continuation.
Therefore, our general torsion conditions derived in this paper, when truncated to the case corresponding to $SU(2)$-structure, agree with the known results from the literature. We regard this as a non-trivial check of the torsion conditions as derived.

In general, it is difficult to find explicit solutions to the above class as the fibration is dependent on the parameter $y$, leading to a Ricci-form $\tilde{R}$ that will also depend on $y$. Motivated by considerations in \cite{jerome1}, one may ask what geometries are permitted when $M_6$ is a complex manifold to see if there is some simple cases.

With that in mind, we focus on  the three-form $\Omega_3 \equiv \Omega \wedge (P^1 + i P^2)$ which defines a natural almost complex structure compatible with $ds^2(M_6)$ and the local $SU(2)$-structure. For $M_6$ complex, we require
\be
d \Omega_3 = V \wedge \Omega_3,
\ee
where $V$ is an arbitrary one-form, but with all other contributions vanishing. It can be shown that the disappearance of the remaining components imposes the further geometric constraint
\be
m C y \cos^2 \zeta \partial_{y} \rho = \frac{1+\sin^2 \zeta}{\cos \zeta} \tilde{J} \cdot d_4 \zeta.
\ee
Compatibility of this condition with  previous (\ref{delyrho}) implies
\be
d_4 \zeta = d_4 \lambda = d_4 A = \partial_{y} \rho = 0.
\ee
Therefore, from (\ref{Om_diff}) we see that $CY_2$ is the only possibility for the base manifold $M_4$ in the case where $M_6$ is complex.

We now recall that the original motivation for this work stemmed from the quest for a back-reacted geometry corresponding to the M5-brane probe with worldvolume $AdS_3 \times H^2 \times S^1$ in the MN geometry. If we insist that the overall metric is invariant under the symmetries of $H^2$, then it appears that (\ref{rescaledOm_diff}) and (\ref{rescaledJ_diff}) are enough to prevent one from constructing such simple solutions with constant curvature Riemann surfaces.

We also remark, that though this class of solutions falls under ``bubbling geometries'', it seems that only the radius of $S^2$ can shrink smoothly at $y=0$. One may determine the boundary conditions at $y=0$  for a smooth
geometry by focusing on the relevant part of the metric: \be
\left(\frac{2y}{m^2 \sin\zeta}\right)^{2/3}
ds^2_{AdS_3}+y^{2/3}\left(\frac{\sin\zeta}{2m}\right)^{4/3}d\Omega^2_2+\frac{dy^2}{e^{4\lambda}\cos^2\zeta}.\ee
Since we also have the relationship, $y = e^{3 \l + A}$, is easy to see that when $y=0$, either the radius of $AdS_3$ or
the radius of $S^2$ vanishes. However, it is impossible to have a
zero-radius $AdS_3$ with the radius of $S^2$ fixed. In fact,
demanding the radius of $S^2$  fixed means $y\sin^2\zeta\equiv t$
fixed. Then $|y|\ge |t|$, so $y$ cannot go to zero.

So, we conclude that only the radius of the $S^2$ shrinks and the radius of $AdS_3$ remains fixed. In this case, the boundary condition for a smooth geometry at $y=0$ is simply
\be \sin\zeta\sim y.\ee

Finally, in this paper we have discussed solutions admitting an internal manifold with both $SU(3)$ and $SU(2)$-structure. In the case of the former, the superconformal symmetry of the dual $\mathcal{N} = (4,0)$ SCFT corresponds to the \textit{small} superconformal symmetry with R-symmetry $SU(2)$ \cite{small}. Despite not having any explicit solutions with $SU(2)$-structure, it is valid to wonder what the symmetry dual to a class of solutions with $SU(2) \times U(1)$ isometry may be. A likely possibility is that the dual theory corresponds to the \textit{large} superconformal symmetry \cite{large} with R-symmetry $SU(2) \times SU(2) \times U(1)$, though one would still have to find another $S^2$ from $M_6$. However, it appears that even when one has an explicit, well-defined geometry, finding the holographic dual is a non-trivial exercise \cite{gukov}. Potentially, a more reasonable source of the R-symmetry may be an LLM parent geometry. In other words, if one can find an explicit solution that interpolates to LLM, then one would be in a position to claim that one has identified the 1/4-BPS back-reacted probe of \cite{Chen:2010jg}.

\section{Discussion}
In this paper we have initiated a classification of the most general $AdS_3 \times S^2$ geometries in M-theory. In contrast to previous work in this area, we have not adopted a restricted ansatz for the fluxes, but instead have assumed a general decomposition of the 11d KSE and have allowed supersymmetry to be our guide. We have subsequently proceeded to write out the general torsion conditions in terms of bilinears constructed from two independent, non-chiral spinors. By demanding consistency among the torsion conditions order by order, we have derived a set of scalar constraints. Although these conditions are also recoverable from the algebraic KSE constraints, surprisingly, the torsion conditions give an extra constraint that may not be deduced from manipulations of the algebraic KSE expressions.

We have also encountered some unexpected surprises along the way. Instead of the anticipated single Killing direction associated with the $U(1)$ R-symmetry, we have identified two Killing directions that generate symmetries of the warped product and field strength ansatz. A third direction which is a Killing direction of the internal $M_6$ fails to generate a symmetry of the $S^2$ warp factor and the two-form flux term $\mathcal{H}$. In fact, if one hopes to incorporate this Killing direction in any internal geometry, then the radius of the $S^2$ will depend on this direction, thus raising lots of questions about the regularity of any class of solutions. For simplicity, we have assumed that $X' = \tilde{K}^1 = 0$ leading to spinors with constant norms. Naturally, relaxing this constraint would be an interesting direction to explore.

Up to this point the analysis had been completely general for an arbitrary internal $M_6$. In the latter part of this paper we have specialised to the case where $M_6$ admits an $SU(2)$-structure defined by two orthogonal chiral spinors, or alternatively two vectors $P^1, P^2$, a real two-form $J$ and a complex two-form $\Omega$. We show from supersymmetry that the four-from flux $\mathcal{G}$ along $M_6$ is necessarily zero and that the assumption that the Killing direction is along only one of $P^1$ or $P^2$ is enough so that the scalar constraints determine the relationship between the two spinors. Using this relationship with $c=0$ to avoid mixing between $J$ and $\Omega$, we integrate the torsion conditions to recover a class of solutions which have previously appeared in \cite{wrappedbranes,nakwoo}. Unfortunately, we are unaware of any explicit solutions in this class. We also remark that setting $\tilde{K}^1 = 0$ essentially rules out the possibility of a non-zero electric flux term $\mathcal{A}$.

When one recalls that the original motivation for this work concerned the hunt for back-reacted M5-brane probes, one is left with some questions. In essence, we have asked if it is possible for a more general flux ansatz to produce a favourable fibration structure allowing for an internal manifold incorporating a hyperbolic space $H^2$. We have derived the general torsion conditions and shown that $\mathcal{G} = 0$ for an $SU(2)$-structure manifold. We have also noted that it is difficult to turn on $\mathcal{A}$, so it appears that the back-reacted probes should correspond to the class of solutions in \cite{nakwoo}. However, it is difficult to find explicit solutions to \cite{nakwoo} and simple attempts to incorporate a hyperbolic space do not work. Indeed, this raises some questions about this class of solutions. As the $U(1)$-fibration structure is decidedly different from LLM, it would be interesting to see if one could find an interpolating solution to establish some connection between these two class of solutions. There are many examples in the literature of solutions that interpolate between $AdS$ vacua in different dimensions.

Our ansatz also makes an assumption about the $S^2$ being round. In general, one may try to squash the sphere to introduce new scalars in analogy with Atiyah-Hitchin manifolds. Indeed, in many recent classifications, round sphere ansatze have featured prominently, but it is a valid question to wonder whether more general solutions can be obtained by squashing the spheres in a manner that also preserves supersymmetry.

In hindsight, we have undertaken the analysis of a difficult class of solutions, with difficulties primarily arising from the freedom of the internal six-dimensional manifold. However, the methods used in deriving the torsion conditions have been quite general and have led to a non-trivial statement about whether the flux term $\mathcal{G}$ may be supported. In practice it should be possible to revisit the simpler LLM set-up, where one only has four internal directions, and show that the flux term ruled out perturbatively can be ruled out in general. One may also be interested in applying the same techniques to find the most general gravity solutions dual to $\mathcal{N}=2$ SCFTs by relaxing the requirement that the two spinors are linearly dependent. Recall from \cite{LLM} that these spinors were taken to be linearly dependent, but in general they may not be.

Finally, with the most general torsion conditions already derived, it may be possible to extend the analysis of this paper to consider the case where the two spinors are linearly independent. We are aware of similar work in \cite{Gauntlett:2005ww} which searched for $AdS_5 \times M_5$ geometries in type IIB outside of the Sasaki-Einstein class of solutions. However, as mentioned, a useful warm up exercise would be tackling this generalisation for the LLM geometries before proceeding to generalise the class of solutions presented here.

\section*{Acknowledgements}
We are grateful to the following for helpful
discussions: Bin Chen, Aristomenis Donos, Jerome Gauntlett, Nakwoo
Kim, Hiroaki Nakajima, Oscar Varela and Piljin Yi. We also benefited from correspondence with Dario Martelli on a late draft.
E\'{O}C would like to extend warm thanks to Imperial College London,
AEI Potsdam and Trinity College Dublin for hospitality during the
early stages of this project. JW would like to thank the Institute of
High Energy Physics and the Chinese Academy of Sciences for
hospitality during a recent trip to Beijing.

\appendix

\section{Torsion conditions}
In this section, we list the torsion conditions for zero, one, two and three-forms. We begin by defining scalars
\bea
W_1 &\equiv& \frac{1}{2} \left( \bar{\e}_{+}^c \e_{+}+\bar{\e}_{-}^c \e_{-}\right) , \quad W_2 \;\equiv\; \frac{1}{2} \left( \bar{\e}_{+}^c \e_{+}-\bar{\e}_{-}^c \e_{-}\right), \nn
X &\equiv& \frac{1}{2} ( \bar{\e}_{+} \e_+ + \bar{\e}_{-} \e_{-} ), \hspace{8mm} \tilde{X} \;\equiv\; \frac{1}{2} ( \bar{\e}_{+} \g_7 \e_+ + \bar{\e}_{-} \g_7 \e_{-} ), \nn
X' &\equiv& \frac{1}{2} (\bar{\e}_{+} \e_+ - \bar{\e}_{-} \e_{-}), \hspace{7mm} \tilde{X}'\; \equiv\; \frac{1}{2} (\bar{\e}_{+} \g_7 \e_+ - \bar{\e}_{-} \g_7 \e_{-}), \nn
Y &\equiv& \bar{\e}_{+} \e_{-},  \hspace{28mm} \tilde{Y} \;\equiv \;\bar{\e}_{+} \g_7 \e_{-}\nn
Z &\equiv& \bar{\e}_{+}^c \e_{-}, \hspace{28mm} \tilde{Z} \; \equiv \; \bar{\e}_{+}^c \g_7 \e_{-},
\eea
vectors
\bea
K^1_m &\equiv& \frac{1}{2} ( \bar{\e}_{+} \g_{m} \e_+ + \bar{\e}_{-} \g_{m} \e_{-} ), \quad \tilde{K}^{1}_m \; \equiv \;   \frac{i}{2} ( \bar{\e}_{+} \g_{m} \g_7 \e_+ + \bar{\e}_{-} \g_{m} \g_7 \e_{-} ), \nn
K^2_m &\equiv& \frac{1}{2} (\bar{\e}_{+} \g_m\e_+ - \bar{\e}_{-} \g_m \e_{-}), \quad \tilde{K}^{2}_m \; \equiv \;  \frac{i}{2} ( \bar{\e}_{+} \g_{m} \g_7 \e_+ - \bar{\e}_{-} \g_{m} \g_7 \e_{-} ),  \nn
K^3_m &\equiv& \bar{\e}_{+} \g_m \e_{-}, \hspace{29mm} \tilde{K}^3_m \; \equiv \; \bar{\e}_{+} \g_m \g_7 \e_{-},\nn
K^4_m &\equiv& \bar{\e}_{+}^c \g_m \e_{-}, \hspace{29mm} \tilde{K}^4_m \; \equiv \;  \bar{\e}_{+}^c \g_m \g_7 \e_{-},
\eea
two-forms denoted by
\bea
L^1_{rm} &\equiv& \frac{i}{2} \left( \bar{\e}_{+} \g_{rm} \e_+ +\bar{\e}_{-}
\g_{rm} \e_-\right) , \quad \tilde{L}^1_{rm} \; \equiv \; \frac{i}{2} \left( \bar{\e}_{+} \g_{rm} \g_7 \e_+ + \bar{\e}_{-}
\g_{rm} \g_7 \e_- \right), \nn
{L}^2_{rm} &\equiv&  \frac{i}{2} \left( \bar{\e}_{+} \g_{rm} \e_+ - \bar{\e}_{-}
\g_{rm} \e_- \right) ,\quad \tilde{L}^2_{rm} \; \equiv \; \frac{i}{2} \left(\bar{\e}_{+} \g_{rm} \g_7
\e_+ - \bar{\e}_{-} \g_{rm} \g_7 \e_- \right), \nn
L^{3}_{rm} &\equiv& \bar{\e}_{+} \g_{rm} \e_{-}, \hspace{32mm} \tilde{L}^{3}_{rm} \;\equiv \;\bar{\e}_{+} \g_{rm} \g_7 \e_{-}, \nn
L^{4}_{rm} &\equiv& \bar{\e}_{+}^{c} \g_{rm} \e_{-},  \hspace{32mm} \tilde{L}^{4}_{rm} \; \equiv \; \bar{\e}^{c}_{+} \g_{rm} \g_7 \e_-, \nn
L^{5}_{rm} &\equiv& \frac{1}{2} \left( \bar{\e}_{+}^{c} \g_{rm} \g_7 \e_{+} +  \bar{\e}_{-}^{c} \g_{rm} \g_7 \e_{-} \right), \nn
L^{6}_{rm} &\equiv& \frac{1}{2} \left( \bar{\e}_{+}^{c} \g_{rm} \g_7 \e_{+} - \bar{\e}_{-}^{c} \g_{rm} \g_7 \e_{-} \right),
\eea
and finally the three-forms
\bea
M^1_{mnp} &\equiv& \frac{i}{2} \left( \bar{\e}_{+} \g_{mnp} \e_{+} +  \bar{\e}_{-} \g_{mnp} \e_{-} \right), \nn
M^2_{mnp} &\equiv& \frac{i}{2} \left( \bar{\e}_{+} \g_{mnp} \e_{+} -  \bar{\e}_{-} \g_{mnp} \e_{-} \right), \nn
M^3_{mnp} &\equiv& \bar{\e}_{+} \g_{mnp} \e_-, \nn
M^4_{mnp} &\equiv& \bar{\e}_+^c \g_{mnp} \e_-, \nn
M^5_{mnp} &\equiv& \frac{1}{2} \left( \bar{\e}_{+}^{c} \g_{mnp} \e_{+} +  \bar{\e}_{-}^{c} \g_{mnp} \e_{-} \right), \nn
M^6_{mnp} &\equiv& \frac{1}{2} \left( \bar{\e}_{+}^{c} \g_{mnp} \e_{+} -  \bar{\e}_{-}^{c} \g_{mnp} \e_{-} \right).
\eea
All other forms can be shown to vanish using the antisymmetric properties of $C_6 \g^{m_1 \cdots m_p}$.

After considerable monotonous, though mildly therapeutic calculation, one may then derive the following scalar torsion conditions:
\bea
\label{seq1} d( e^{3 \l + A} W_1) &=& 0, \\
\label{seq2} e^{-3 \l} d(e^{3 \l} W_2) &=& -2 m \tilde{K}^{4} - \tilde{Z} e^{-3 \l} \mathcal{A}, \\
\label{seq3} d X &=& 0, \\
\label{seq4} e^{-3 \l} d (e^{3 \l} \tilde{X}) &=& 2 m \Re(K^3) - \Re(Y) e^{-3 \l} \mathcal{A}, \\
\label{seq5} e^{2 A} d (e^{-A} X') &=& \tilde{K}^1, \\
\label{seq6}d (e^{3 \lambda + A} \tilde{X}') &=& 0, \\
\label{seq7}d (\Re(Y)) &=& 0, \\
\label{seq8}d (e^{-A} \Im(Y)) &=& 0, \\
\label{seq9} e^{-3 \l} d(e^{3 \l} \Re(\tilde{Y})) &=& 2 m K^1 - X e^{-3 \l} \mathcal{A}, \\
\label{seq10} e^{-3 \l} d(e^{3 \l +A} \Im (\tilde{Y})) &=& - \Re(K^3), \\
\label{seq11}e^{-3 \l } d ( e^{3 \l + A} Z) &=& i \tilde{K}^4, \\
\label{seq12}d \tilde{Z} &=& 0,
\eea
vector torsion conditions:
\bea
\label{veq1} d ( e^{3\l} K^1) &=& 0, \\
\label{veq2} d(e^{6 \l +2 A} \tilde{K}^{1}) &=& e^{3 \l} \Re(\tilde{Y}) {\cal H } + 2 m e^{6 \l +2 A} \Im(L^3) - e^{3 \l + 2 A} X *_6 {\cal G}, \\
\label{veq3} d(e^{3 \l +A} K^2) &=& - e^{-A} \Im(Y) {\cal H }  + e^{3 \l } \tilde{L}^{1}, \\
\label{veq3b} d (e^{6 \l +A} \tilde{K}^2) &=& e^{3 \l +A} {\cal A} \wedge \Im(K^3) - e^{3 \l + A} X' *_6 {\cal G}, \\
\label{veq4} d(e^{3 \l +A} \Im(K^3) ) &=& e^{-A} X' {\cal H }, \\
\label{veq5} d(e^{3 \l} \Re(K^3)) &=& 0, \\
\label{veq6} d(e^{6 \l +A} \Re(\tilde{K}^3)) &=& - e^{3 \l + A} \Im(Y) *_6 {\cal G} - e^{3 \l +A} \mathcal{A} \wedge K^2 + e^{6 \l} \Im (L^3), \\
\label{veq7} d(e^{6 \l + 2A } \Im(\tilde{K}^{3})) &=& - e^{3 \l} \tilde{X} {\cal H } + 2 m e^{6 \l + 2 A} L^1 + e^{3 \l + 2 A} \Re(Y) *_{6} {\cal G} \nn &+& 2 e^{6 \l +A} \Re(L^3), \\
\label{veq8}d(e^{6 \l + 2 A} K^4) &=& i W^2 e^{3 \l} {\cal H } + 2 m e^{6 \l + 2 A} L^6 - i \tilde{Z} e^{3 \l + 2 A} *_6 {\cal G} - 2 i e^{6 \l + A} \tilde{L}^4, \\
\label{veq9} d(e^{3 \l} \tilde{K}^{4}) &=& 0.
 \eea
two-form torsion conditions \footnote{$A \lrcorner B = \frac{1}{q!} A^{\a^1 \cdots \a^{q}} B^{\a^1 \cdots \a^q \b^1 \cdots \b^p}$. }:
\bea
\label{2eq1} d (e^{6 \l + 2 A} L^1 ) &=&  e^{3 \l} \Re(K^3) \wedge {\cal H } -2 e^{6 \l +A} *_6 M^2, \\
\label{2eq2} d(e^{6 \l +A} L^2) &=& - 2 m e^{6 \l +A} *_6 \Re(M^3) -  e^{6 \l} *_6 M^1, \\
\label{2eq3} d (e^{3 \l} \tilde{L}^{1} ) &=& - \tilde{K}^{1} \lrcorner {\cal G}, \\
\label{2eq4} d(e^{3 \l +A } \tilde{L}^2 ) &=& - e^{-A} \Re(\tilde{K}^3)  \wedge {\cal H } - e^{A} \tilde{K}^{1} \lrcorner {\cal G} - e^{A} *_6 (\mathcal{A} \wedge \Re(\tilde{L}^3)), \\
\label{2eq5}d (e^{6 \l + 2 A} \Im(L^3)) &=& - e^{3 \l} K^1 \wedge {\cal H }, \\
\label{2eq6} d(e^{6 \l + A} \Re(L^3)) &=& 2 m e^{6 \l +A} *_6 M^2,\\
\label{2eq7}d(e^{3 \l} \Im(\tilde{L}^{3}) ) &=& - \Im(\tilde{K}^3) \lrcorner {\cal G}, \\
\label{2eq8} d(e^{3 \l +A} \Re(\tilde{L}^3)) &=& e^{-A} \tilde{K}^2 \wedge {\cal H } - e^{A} \Re(\tilde{K}^3) \lrcorner {\cal G} + e^{3 \l} \Im(M^3) \nn &+& e^{A} *_6 ( {\cal A} \wedge \tilde{L}^2 ), \\
\label{2eq9}d(e^{6 \l + A} L^5 ) &=& 2 m e^{6 \l + A} M^4 + i e^{6 \l} M^6, \\
\label{2eq10}d (e^{6 \l + 2 A} L^6 ) &=& i e^{3 \l} \tilde{K}^4 \wedge {\cal H } + 2 i e^{6 \l + A} M^5, \\
\label{2eq11} d(e^{3 \l} L^4 ) &=& K^4 \lrcorner {\cal G}, \\
\label{2eq12} d(e^{6 \l +A} \tilde{L}^{4}) &=& 2 m e^{6 \l +A} M^5,
\eea
and finally three-form torsion conditions:
\bea
\label{3eq1} d(e^{6 \l +A}M^5) &=& - e^{3 \l +A} W^1 {\cal G}, \\
\label{3e2} d(e^{6 \l +A} *_6 M^2) &=& - e^{3 \l +A} \tilde{X}' {\cal G}, \\
\label{3eq3} d(e^{6 \l} M^6) &=&  2 m i e^{6 \l} *_6 L^4 - e^{3 \l} W^2 {\cal G} + e^{3 \l - 2 A} *_6 {\cal H } \tilde{Z}, \\
\label{3eq4} d(e^{6 \l + A} M^4 ) &=& -e^{3 \l +A} Z {\cal G} + e^{6 \l} *_6 L^4, \\
\label{3eq5} d (e^{6 \l} *_6 M^1) &=& - \tilde{X} e^{3 \l} {\cal G} + e^{3 \l -2 A} \Re(Y) *_6 {\cal H } - 2 m e^{6 \l} *_6 \Im(\tilde{L}^3), \\
\label{3eq6} d(e^{6 \l +A} *_6 \Re(M^3)) &=& i e^{3 \l +A} \tilde{X}' {\cal G} + e^{6 \l} *_6 \Im(\tilde{L}^3).
\eea
Throughout we have used $\Re$ and $\Im$ to denote the real and imaginary part of expressions

\section{$SU(2)$-structure Killing spinors}
In this appendix, we translate the scalar constraints (\ref{genconst}) on the geometry into relationships amongst the complex functions $a_i$ appearing in the definition of $\e_-$ (\ref{2spinors}). These constraints then take the form:
\bea
W_1 &=& \bar{\e}_+^c \e_+ + \bar{\e}^c \e_- =  c \cos \zeta \left( 1+ a_1^2 -a_2^2 \right) + c^* \cos \zeta \left(a_3^2 - a_4^2\right) \nn &+& 2 (a_1 a_3-a_2 a_4) + 2 \sin \zeta (a_1 a_4 -a_2 a_3) = 0, \nn
\tilde{Z} &=& a_2 c \cos \zeta + a_3 \sin \zeta + a_4 = 0, \nn
\Re(Y) &=& \Re (a_1) - \Re(a_2) \sin \zeta + \Re(a_3 c^*) \cos \zeta = 0, \nn
\bar{\e}_- \e_- &=& \sum |a_i|^2 + 2 \sin \zeta \left[ -  \Re(a_1 a_2^*) + \Re(a_3 a_4^*) \right]  \nn &+& 2 \cos \zeta \left[ \Re(a_1 a_3^* c) + \Re(a_2 a_4^* c) \right] = 1 , \nn
\tilde{X}' &=& \frac{1}{2} ( \bar{\e}_+ \g_7 \e_+ - \bar{\e}_- \g_7 \e_-), \nn
&=& - \frac{1}{2} \sin \zeta (1 - |a_1|^2 - |a_2|^2 + |a_3|^2 + |a_4|^2 ) -  \Re(a_1 a_2^* + a_3 a_4^*) \nn &-& \cos \zeta \Re(a_1 a_4^* c + a_2 a_3^* c) = 0, \nn
 i e^{-A} c \cos \zeta  &=& - 2 m \left( a_1 c \cos \zeta + a_3 + a_4 \sin \zeta \right), \nn
e^{-A} \sin \zeta &=& 2 m \left( - \Im(a_1) \sin \zeta + \Im(a_2) + \Im(a_4 c^* ) \cos \zeta \right). \label{scalarcond1}
\eea
Note, we have fewer constraints than functions, so it is not possible to solve these equations in a closed form. We will proceed by making assumptions about the Killing direction. With $SU(2)$-structure, one may define two vectors $V_i, i =1,2$, in terms of which the other vectors may be written,
\bea
K^1 &=& \frac{1}{2} \left(1 + |a_1|^2 + |a_4|^2 - |a_2|^2 - |a_3|^2 \right) V_1 +\Im(a_1 a_2^* - a_3 a_4^*) V_2, \nn
\tilde{K}^1 &=& \Im(a_1 a_2^* - a_3 a_4^*) V_1 - \frac{1}{2} \left(1 + |a_1|^2 + |a_4|^2 - |a_2|^2 - |a_3|^2 \right) V_2 , \nn
K^2 &=& \frac{1}{2} \left(1 - |a_1|^2 - |a_4|^2 + |a_2|^2 + |a_3|^2 \right) V_1 - \Im(a_1 a_2^* - a_3 a_4^*) V_2, \nn
\tilde{K}^2 &=& -\Im(a_1 a_2^* - a_3 a_4^*) V_1 - \frac{1}{2} \left(1 - |a_1|^2 - |a_4|^2 + |a_2|^2 + |a_3|^2 \right) V_2, \nn
K^3 &=& a_1 V_1 + a_2 i V_2, \nn
i \Im(\tilde{K}^3) &=& \Re(a_1) i V_2 + i \Im(a_2) V_1, \nn
\Re(\tilde{K}^3) &=& - \Im(a_1) V_2 + \Re(a_2) V_1, \nn
K^4 &=& -a_3 V_1 - a_4 i V_2, \nn
\tilde{K}^4 &=& - a_3 i V_2 - a_4 V_1,
\eea
where we have defined
\be
V^m_1 = \bar{\e}_+ \g^{m} \e_+, \quad V_2^{m} = i \bar{\e}_+ \g^{m} \g_7 \e_+.
\ee
We also note that by imposing $X'= \tilde{K}^1 = 0$, in addition we also have
\bea
\label{scalarcond2}
\Im (a_1 a_2^*) &=& \Im(a_3 a_4^*), \nn
1 + |a_1|^2 + |a_4|^2 &=& |a_2|^2 + |a_3|^2,
\eea
with necessarily $K^1 = 0$.

Henceforth, we will make the simplifying assumption that \textit{either} $V_1$ \textit{or} $V_2$ defines a Killing direction. In other words, as both $\Im(\tilde{K}^3)$ and $K^4$ are Killing we will take them to be proportional, and will either set $ \Re(a_1) = a_4 = 0$ or $\Im(a_2) = a_3 = 0$ from the offset so that the Killing direction is respectively along either $V_1$ or $V_2$.

If we assume that the Killing direction is along $V_1$ i.e. $\Re(a_1) = a_4 = 0$, then one can show from the above constraints that the only solution is \be
a_2 = i \frac{ e^{-A}}{2m} \sin \zeta, \quad a_3 = - i \frac{e^{-A} }{2m} c \cos \zeta, \quad \Im(a_1) = 0.
\ee
This leads to the form of the general spinor (\ref{genspinor}) quoted in the text. However, if one tries to adopt $V_2$ as the Killing direction by imposing $\Im(a_2) = a_3 = 0$, then the scalar constraints permit no solution.


\begin{thebibliography}{99}


\bibitem{Gaiotto:2009we}
  D.~Gaiotto,
  ``N=2 dualities,''
  arXiv:0904.2715 [hep-th].

\bibitem{AGT}
  L.~F.~Alday, D.~Gaiotto and Y.~Tachikawa,
  ``Liouville Correlation Functions from Four-dimensional Gauge Theories,''
  Lett.\ Math.\ Phys.\  {\bf 91} (2010) 167
  [arXiv:0906.3219 [hep-th]].

\bibitem{nonlocalops}
  N.~Drukker, D.~R.~Morrison and T.~Okuda,
  ``Loop operators and S-duality from curves on Riemann surfaces,''
  JHEP {\bf 0909}, 031 (2009)
  [arXiv:0907.2593 [hep-th]],
  L.~F.~Alday, D.~Gaiotto, S.~Gukov, Y.~Tachikawa and H.~Verlinde,
  ``Loop and surface operators in N=2 gauge theory and Liouville modular
  geometry,''
  arXiv:0909.0945 [hep-th],
  N.~Drukker, J.~Gomis, T.~Okuda and J.~Teschner,
  ``Gauge Theory Loop Operators and Liouville Theory,''
  arXiv:0909.1105 [hep-th],
  D.~Gaiotto,
  ``Surface Operators in N=2 4d Gauge Theories,''
  arXiv:0911.1316 [hep-th],
  J.~F.~Wu and Y.~Zhou,
  ``From Liouville to Chern-Simons, Alternative Realization of Wilson Loop
  Operators in AGT Duality,''
  arXiv:0911.1922 [hep-th].

\bibitem{Chen:2010jg}
  B.~Chen, E.~O.~Colgain, J.~B.~Wu and H.~Yavartanoo,
  ``N = 2 SCFTs: An M5-brane perspective,''
  arXiv:1001.0906 [hep-th].

\bibitem{LLM}
  H.~Lin, O.~Lunin and J.~M.~Maldacena,
  ``Bubbling AdS space and 1/2 BPS geometries,''
  JHEP {\bf 0410} (2004) 025
  [arXiv:hep-th/0409174].

\bibitem{Donos1}
A.~Donos,
  ``A description of 1/4 BPS configurations in minimal type IIB SUGRA,''
  Phys.\ Rev.\  D {\bf 75}, 025010 (2007)
  [arXiv:hep-th/0606199],

\bibitem{eoin}
 P.~Figueras, O.~A.~P.~Mac Conamhna and E.~O Colgain,
  ``Global geometry of the supersymmetric AdS(3)/CFT(2) correspondence in
  M-theory,''
  Phys.\ Rev.\  D {\bf 76}, 046007 (2007)
  [arXiv:hep-th/0703275],

\bibitem{bubbles}
  A.~Donos,
  ``BPS states in type IIB SUGRA with SO(4) x SO(2)(gauged) symmetry,''
  JHEP {\bf 0705} (2007) 072
  [arXiv:hep-th/0610259],
  N.~Kim,
  ``AdS(3) solutions of IIB supergravity from D3-branes,''
  JHEP {\bf 0601} (2006) 094
  [arXiv:hep-th/0511029],
  E.~Gava, G.~Milanesi, K.~S.~Narain and M.~O'Loughlin,
  ``1/8 BPS states in AdS/CFT,''
  JHEP {\bf 0705}, 030 (2007)
  [arXiv:hep-th/0611065],
  O.~A.~P.~Mac Conamhna and E.~O Colgain,
  ``Supersymmetric wrapped membranes, AdS(2) spaces, and bubbling geometries,''
  JHEP {\bf 0703} (2007) 115
  [arXiv:hep-th/0612196],
  N.~Kim and J.~D.~Park,
  ``Comments on AdS(2) solutions of D = 11 supergravity,''
  JHEP {\bf 0609} (2006) 041
  [arXiv:hep-th/0607093],
  J.~P.~Gauntlett, N.~Kim and D.~Waldram,
  ``Supersymmetric AdS(3), AdS(2) and bubble solutions,''
  JHEP {\bf 0704} (2007) 005
  [arXiv:hep-th/0612253],
   J.~P.~Gauntlett and O.~A.~P.~Mac Conamhna,
  ``AdS spacetimes from wrapped D3-branes,''
  Class.\ Quant.\ Grav.\  {\bf 24} (2007) 6267
  [arXiv:0707.3105 [hep-th]].

\bibitem{Chen2}
  B.~Chen {\it et al.},
  ``Bubbling AdS and droplet descriptions of BPS geometries in IIB
  supergravity,''
  JHEP {\bf 0710} (2007) 003
  [arXiv:0704.2233 [hep-th]].



\bibitem{MG}
  D.~Gaiotto and J.~Maldacena,
  ``The gravity duals of N=2 superconformal field theories,''
  arXiv:0904.4466 [hep-th].



\bibitem{MN}
  J.~M.~Maldacena and C.~Nunez,
  ``Supergravity description of field theories on curved manifolds and a no  go
  theorem,''
  Int.\ J.\ Mod.\ Phys.\  A {\bf 16}, 822 (2001)
  [arXiv:hep-th/0007018].

\bibitem{backx}
E.~D'Hoker, J.~Estes and M.~Gutperle,
  ``Gravity duals of half-BPS Wilson loops,''
  JHEP {\bf 0706} (2007) 063
  [arXiv:0705.1004 [hep-th]],
   O.~Lunin,
  ``On gravitational description of Wilson lines,''
  JHEP {\bf 0606}, 026 (2006)
  [arXiv:hep-th/0604133],
   S.~Yamaguchi,
  ``Bubbling geometries for half BPS Wilson lines,''
  Int.\ J.\ Mod.\ Phys.\  A {\bf 22}, 1353 (2007)
  [arXiv:hep-th/0601089],
O.~Lunin,
  ``1/2-BPS states in M theory and defects in the dual CFTs,''
  JHEP {\bf 0710} (2007) 014
  [arXiv:0704.3442 [hep-th]],
    E.~D'Hoker, J.~Estes, M.~Gutperle and D.~Krym,
  ``Exact Half-BPS Flux Solutions in M-theory I, Local Solutions,''
  JHEP {\bf 0808}, 028 (2008)
  [arXiv:0806.0605 [hep-th]],
  E.~D'Hoker, J.~Estes, M.~Gutperle and D.~Krym,
  ``Exact Half-BPS Flux Solutions in M-theory II: Global solutions asymptotic
  to $AdS_7 \times S^4$,''
  JHEP {\bf 0812}, 044 (2008)
  [arXiv:0810.4647 [hep-th]],
J.~Gomis and S.~Matsuura,
  ``Bubbling Surface Operators And S-Duality,''
  JHEP {\bf 0706}, 025 (2007)
  [arXiv:0704.1657 [hep-th]].

\bibitem{jerome1}
  J.~P.~Gauntlett, D.~Martelli, J.~Sparks and D.~Waldram,
  ``Supersymmetric AdS(5) solutions of M-theory,''
  Class.\ Quant.\ Grav.\  {\bf 21}, 4335 (2004)
  [arXiv:hep-th/0402153].

\bibitem{Gstructure}
  J.~P.~Gauntlett, D.~Martelli, S.~Pakis and D.~Waldram,
  ``G-structures and wrapped NS5-branes,''
  Commun.\ Math.\ Phys.\  {\bf 247}, 421 (2004)
  [arXiv:hep-th/0205050].

\bibitem{d11kill}
  J.~P.~Gauntlett and S.~Pakis,
  ``The geometry of D = 11 Killing spinors,''
  JHEP {\bf 0304} (2003) 039
  [arXiv:hep-th/0212008],
  J.~P.~Gauntlett, J.~B.~Gutowski and S.~Pakis,
  ``The geometry of D = 11 null Killing spinors,''
  JHEP {\bf 0312} (2003) 049
  [arXiv:hep-th/0311112].

 \bibitem{oisin}
  O.~A.~P.~Mac Conamhna,
  ``Eight-manifolds with G-structure in eleven dimensional supergravity,''
  Phys.\ Rev.\  D {\bf 72} (2005) 086007
  [arXiv:hep-th/0504028],
  M.~Cariglia and O.~A.~P.~Mac Conamhna,
  ``Null structure groups in eleven dimensions,''
  Phys.\ Rev.\  D {\bf 73} (2006) 045011
  [arXiv:hep-th/0411079],  
  
\bibitem{nakwoo}
  H.~Kim, K.~K.~Kim and N.~Kim,
  ``1/4-BPS M-theory bubbles with SO(3) x SO(4) symmetry,''
  JHEP {\bf 0708}, 050 (2007)
  [arXiv:0706.2042 [hep-th]].



\bibitem{wrappedbranes}
  J.~P.~Gauntlett, O.~A.~P.~Mac Conamhna, T.~Mateos and D.~Waldram,
  ``AdS spacetimes from wrapped M5 branes,''
  JHEP {\bf 0611} (2006) 053
  [arXiv:hep-th/0605146].

\bibitem{MSW}
  J.~M.~Maldacena, A.~Strominger and E.~Witten,
  ``Black hole entropy in M-theory,''
  JHEP {\bf 9712}, 002 (1997)
  [arXiv:hep-th/9711053].

\bibitem{romans}
  L.~J.~Romans,
  ``Gauged N=4 Supergravities In Five-Dimensions And Their Magnetovac
  Backgrounds,''
  Nucl.\ Phys.\  B {\bf 267} (1986) 433.

\bibitem{oscar}
  J.~P.~Gauntlett and O.~Varela,
  ``D=5 SU(2)xU(1) Gauged Supergravity from D=11 Supergravity,''
  JHEP {\bf 0802}, 083 (2008)
  [arXiv:0712.3560 [hep-th]].


\bibitem{jerome5}
  J.~P.~Gauntlett, D.~Martelli and D.~Waldram,
  ``Superstrings with intrinsic torsion,''
  Phys.\ Rev.\  D {\bf 69}, 086002 (2004)
  [arXiv:hep-th/0302158].


\bibitem{dario}
  D.~Martelli and J.~Sparks,
  ``G-structures, fluxes and calibrations in M-theory,''
  Phys.\ Rev.\  D {\bf 68}, 085014 (2003)
  [arXiv:hep-th/0306225].



\bibitem{Gauntlett:2005ww}
  J.~P.~Gauntlett, D.~Martelli, J.~Sparks and D.~Waldram,
  ``Supersymmetric AdS(5) solutions of type IIB supergravity,''
  Class.\ Quant.\ Grav.\  {\bf 23}, 4693 (2006)
  [arXiv:hep-th/0510125].

\bibitem{pope1}
  H.~Lu, C.~N.~Pope and J.~Rahmfeld,
  ``A construction of Killing spinors on S**n,''
  J.\ Math.\ Phys.\  {\bf 40}, 4518 (1999)
  [arXiv:hep-th/9805151].

\bibitem{sohnius}
  M.~F.~Sohnius,
  ``Introducing Supersymmetry,''
  Phys.\ Rept.\  {\bf 128} (1985) 39.

\bibitem{small}
  M.~Ademollo {\it et al.},
  ``Dual String With U(1) Color Symmetry,''
  Nucl.\ Phys.\  B {\bf 111}, 77 (1976), 
  M.~Ademollo {\it et al.},
  ``Supersymmetric Strings And Color Confinement,''
  Phys.\ Lett.\  B {\bf 62}, 105 (1976).
    
\bibitem{large}
  A.~Sevrin, W.~Troost and A.~Van Proeyen,
  ``Superconformal Algebras in Two-Dimensions with N=4,''
  Phys.\ Lett.\  B {\bf 208}, 447 (1988).
  

\bibitem{gukov}
  S.~Gukov, E.~Martinec, G.~W.~Moore and A.~Strominger,
  ``The search for a holographic dual to AdS(3) x S**3 x S**3 x S**1,''
  Adv.\ Theor.\ Math.\ Phys.\  {\bf 9}, 435 (2005)
  [arXiv:hep-th/0403090].
\end{thebibliography}
\end{document}